\documentclass[conference]{IEEEtran}
\IEEEoverridecommandlockouts
\usepackage{orcidlink}
\usepackage{amsmath}
\usepackage{graphicx}
\usepackage{float}
\usepackage{enumitem}
\usepackage{amssymb}
\usepackage{amsfonts}
\usepackage{booktabs}
\usepackage{tabularx}
\usepackage{multirow}
\usepackage[caption=false,font=footnotesize]{subfig}

\usepackage{tikz}
\usepackage{xcolor}
\usepackage{listings}
\definecolor{codekeyword}{rgb}{0.00,0.00,0.70}
\definecolor{codecomment}{rgb}{0.00,0.45,0.00}
\definecolor{codestring}{rgb}{0.65,0.10,0.10}
\usepackage{hyperref}
\usepackage[ruled,vlined,linesnumbered]{algorithm2e}
\usepackage{microtype}

\usetikzlibrary{positioning,arrows.meta,fit,backgrounds,calc,shapes.geometric,matrix,decorations.pathreplacing}
\graphicspath{{images/}}

\DeclareRobustCommand{\NumaTorch}{\mbox{NUMA-Torch}}

\newif\ifreview
\reviewfalse
\ifreview
  \usepackage[switch,mathlines]{lineno}
  \newcommand*\patchAmsMathEnvironmentForLineno[1]{%
    \expandafter\let\csname old#1\expandafter\endcsname\csname #1\endcsname
    \expandafter\let\csname oldend#1\expandafter\endcsname\csname end#1\endcsname
    \renewenvironment{#1}%
      {\linenomath\csname old#1\endcsname}%
      {\csname oldend#1\endcsname\endlinenomath}%
  }
  \newcommand*\patchBothAmsMathEnvironmentsForLineno[1]{%
    \patchAmsMathEnvironmentForLineno{#1}%
    \patchAmsMathEnvironmentForLineno{#1*}%
  }
  \AtBeginDocument{%
    \patchBothAmsMathEnvironmentsForLineno{equation}%
    \patchBothAmsMathEnvironmentsForLineno{align}%
    \patchBothAmsMathEnvironmentsForLineno{flalign}%
    \patchBothAmsMathEnvironmentsForLineno{alignat}%
    \patchBothAmsMathEnvironmentsForLineno{gather}%
    \patchBothAmsMathEnvironmentsForLineno{multline}%
  }
\fi

\title{\NumaTorch{}: Compiler-Driven NUMA-Aware LLM Prefill on Multi-Socket CPUs}

\author{%
\IEEEauthorblockN{%
Thisara Rathnayaka Weerakoon\orcidlink{0009-0002-1961-2659}\IEEEauthorrefmark{1},
Ishan Dewpura\orcidlink{0009-0005-2629-9316}\IEEEauthorrefmark{1},
Niman Ransindu\orcidlink{0009-0009-5439-3317}\IEEEauthorrefmark{1},\\
Shavin Anjitha\orcidlink{0009-0003-1535-0898}\IEEEauthorrefmark{1},
Tharindu R. Patabandi\orcidlink{0000-0002-5052-8183}\IEEEauthorrefmark{2},
Sanath Jayasena\orcidlink{0000-0001-5097-8769}\IEEEauthorrefmark{1}}
\IEEEauthorblockA{\IEEEauthorrefmark{1}Dept.\ of Computer Science \& Engineering, University of Moratuwa, Sri Lanka\\
\{thisara.21, ishandewpura.21, niman.21, shavin.21, sanath\}@cse.mrt.ac.lk}
\IEEEauthorblockA{\IEEEauthorrefmark{2}Independent Researcher, San Jose, CA, USA\\
tharindu@protonmail.com}
\thanks{\textcopyright{}~2026 IEEE. Personal use of this material is permitted. Permission from IEEE must be obtained for all other uses, in any current or future media, including reprinting/republishing this material for advertising or promotional purposes, creating new collective works, for resale or redistribution to servers or lists, or reuse of any copyrighted component of this work in other works. Accepted for publication in the 2026 IEEE 33rd International Conference on High Performance Computing, Data, and Analytics (HiPC), Bengaluru, India, December 16--19, 2026. This is the authors' accepted version; the final published version will appear in IEEE Xplore.}}

\begin{document}

\bstctlcite{IEEEexample:BSTcontrol}

\raggedbottom

\ifreview\linenumbers\fi

\maketitle

\begin{abstract}
For sufficiently large transformer-based Large Language Models (LLMs), whose parameters cannot be retained in on-chip caches, prefill throughput on multi-socket CPUs is often limited by sustainable DRAM bandwidth rather than peak floating-point throughput. In the standard PyTorch compilation (i.e., TorchInductor) path, model weights are allocated without considering the Non-Uniform Memory Access (NUMA) topology—specifically, the affinity between memory banks and compute cores is disregarded. This causes a substantial fraction of weight accesses to traverse the inter-socket interconnect, severely degrading the effective memory bandwidth available to each socket. We address this bottleneck through a compiler-level solution that makes TorchInductor NUMA-aware. Treating each NUMA socket as an independent tensor-parallel partition, we introduce NUMA-aware Tensor Parallelism (NUMA-TP). This approach transparently shards weights across sockets and schedules compute kernels such that thread pools and memory allocations remain strictly local to their respective sockets using CPU affinity. We evaluate the approach on GPT-2~\cite{radford2019gpt2} (117M, 345M, 762M) and BERT~\cite{devlin2019bert} (Small, Large), which exercise standard multi-head attention and MLP, as well as Gemma~3 270M~\cite{team2024gemma}, which combines Multi-Query Attention with a GeGLU-gated MLP. By strictly enforcing data locality, our approach successfully mitigates cross-socket traffic, achieving prefill speedups of up to $2.63\times$ on a dual-socket Haswell platform, up to $1.87\times$ on a dual-socket Cascade Lake platform, and up to $2.27\times$ on a four-socket Haswell-EX platform.
\end{abstract}


\section{Introduction}
\label{sec:introduction}

Driven by the necessity of accommodating many cores, modern
high-performance servers have universally adopted Non-Uniform Memory
Access (NUMA) architectures~\cite{calciu2017numa}. There is consequently a tremendous, yet underutilized, potential to deploy Large Language Model (LLM) inference on these multi-socket CPU platforms, given their cost-effectiveness, broad availability, and the frequent absence of GPUs in deployment environments. The \textit{prefill} phase of LLM inference which processes the full input prompt in one forward pass is traditionally assumed compute-bound owing to the high arithmetic intensity of dense matrix multiplications over long sequences. Yet on multi-socket CPUs under the standard TorchInductor compilation pipeline, prefill unexpectedly becomes severely DRAM-bound as NUMA-oblivious memory allocations force massive cross-socket traffic.

\begin{figure}[t]
\centering
\resizebox{\linewidth}{!}{%
\begin{tikzpicture}[
    font=\small,
    >={Latex[length=2mm]},
    dram/.style={draw,thick,fill=blue!10,minimum width=1.65cm,minimum height=0.65cm,rounded corners=2pt,align=center},
    socket/.style={draw,thick,dashed,rounded corners=4pt,inner sep=4pt,fill=gray!5},
    core/.style={draw,thick,fill=orange!20,minimum width=1.3cm,minimum height=0.58cm,rounded corners=2pt},
    cache/.style={draw,thick,fill=green!15,minimum width=1.3cm,minimum height=0.45cm,rounded corners=2pt},
    l3/.style={draw,thick,fill=yellow!25,minimum width=3.35cm,minimum height=0.52cm,rounded corners=2pt},
    mc/.style={draw,thick,fill=purple!15,minimum width=1.65cm,minimum height=0.42cm,rounded corners=2pt},
    qpi/.style={<->,very thick,red!70!black,line width=1.2pt}
]

\node[core] (c00) at (0,0) {Core 0};
\node[core,right=0.25cm of c00] (c01) {Core 1};
\node[cache,below=0.1cm of c00] (l1_00) {L1\,/\,L2};
\node[cache,below=0.1cm of c01] (l1_01) {L1\,/\,L2};

\node[l3,below=0.25cm of $(l1_00)!0.5!(l1_01)$] (l3_0) {Shared L3 Cache};

\node[mc,below=0.08cm of l3_0] (mc0) {Mem.\ Controller};
\begin{scope}[on background layer]
  \node[socket,fit=(c00)(c01)(l1_00)(l1_01)(l3_0)(mc0),label={[font=\small\bfseries,fill=white,inner sep=1pt]above:Socket 0\,/\,NUMA Node 0}] (sock0) {};
\end{scope}
\node[dram,left=0.25cm of sock0] (dram0) {DRAM 0\\(local)};
\draw[thick,<->] (dram0.east) -- ++(0.12,0) |- (mc0.west);

\node[core,right=2.15cm of c01] (c10) {Core 0};
\node[core,right=0.25cm of c10] (c11) {Core 1};
\node[cache,below=0.1cm of c10] (l1_10) {L1\,/\,L2};
\node[cache,below=0.1cm of c11] (l1_11) {L1\,/\,L2};

\node[l3,below=0.25cm of $(l1_10)!0.5!(l1_11)$] (l3_1) {Shared L3 Cache};

\node[mc,below=0.08cm of l3_1] (mc1) {Mem.\ Controller};
\begin{scope}[on background layer]
  \node[socket,fit=(c10)(c11)(l1_10)(l1_11)(l3_1)(mc1),label={[font=\small\bfseries,fill=white,inner sep=1pt]above:Socket 1\,/\,NUMA Node 1}] (sock1) {};
\end{scope}
\node[dram,right=0.25cm of sock1] (dram1) {DRAM 1\\(local)};
\draw[thick,<->] (dram1.west) -- ++(-0.12,0) |- (mc1.east);

\draw[thick] (c00.south) -- (l1_00.north);
\draw[thick] (c01.south) -- (l1_01.north);
\draw[thick] (l1_00.south) -- (l1_00.south|-l3_0.north);
\draw[thick] (l1_01.south) -- (l1_01.south|-l3_0.north);
\draw[thick] (l3_0.south) -- (mc0.north);

\draw[thick] (c10.south) -- (l1_10.north);
\draw[thick] (c11.south) -- (l1_11.north);
\draw[thick] (l1_10.south) -- (l1_10.south|-l3_1.north);
\draw[thick] (l1_11.south) -- (l1_11.south|-l3_1.north);
\draw[thick] (l3_1.south) -- (mc1.north);

\draw[qpi] (sock0.east) -- node[above=1pt,font=\small\bfseries,text=red!60!black,fill=white,inner sep=1pt]{QPI\,/\,UPI} (sock1.west);
\node[font=\small\itshape,text=red!60!black,align=center,fill=white,inner sep=1pt,below=5pt of $(sock0.east)!0.5!(sock1.west)$] {cross-socket\\interconnect};

\node[font=\footnotesize\itshape,below=0.05cm of dram0] {local access};
\node[font=\footnotesize\itshape,below=0.05cm of dram1] {local access};

\draw[->,dashed,thick,red!50!black]
    (c00.north) to[bend left=35]
    node[above,font=\footnotesize\itshape,text=red!50!black]{remote DRAM access (high cost)}
    (dram1.north);

\end{tikzpicture}%
}

\caption{Logical view of a two-socket NUMA CPU treated as a distributed
system; the dashed arrow marks a high-cost remote DRAM access across the
cross-socket interconnect.}
\label{fig:numa-topology}
\end{figure}
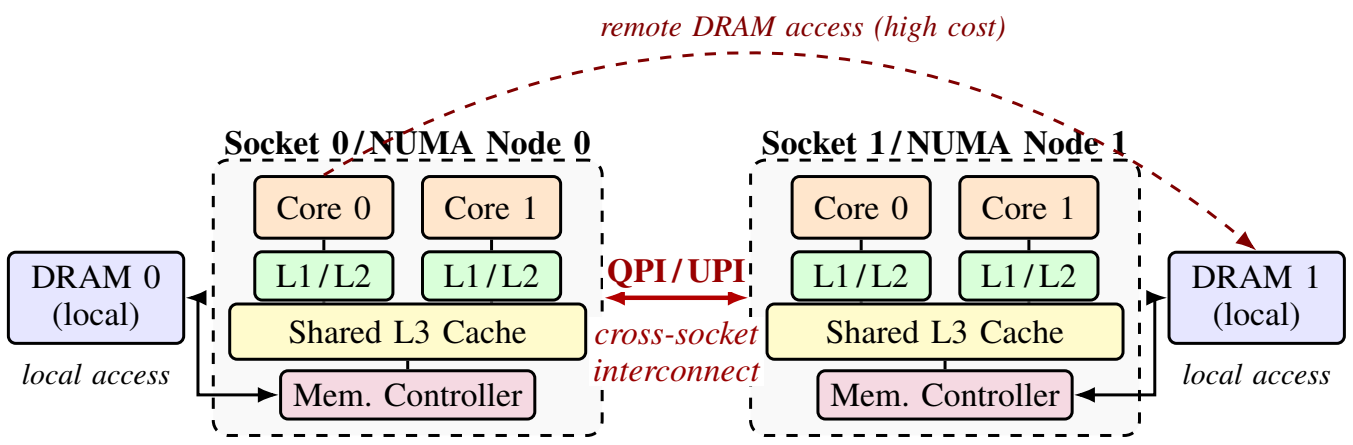

As illustrated in Figure~\ref{fig:numa-topology}, we view each socket of a
multi-socket NUMA CPU as an independent node owning its locally attached DRAM
and memory controller, together with cores that have private L1/L2 caches and
share an L3. Sockets are linked by a QPI/UPI interconnect serving as the
inter-rank network; accesses to another socket's memory must cross it,
incurring substantially higher latency and lower bandwidth than local accesses,
and are therefore treated as explicit communication in our compilation
model.

\begin{figure}[t]
    \centering
    \includegraphics[width=\linewidth]{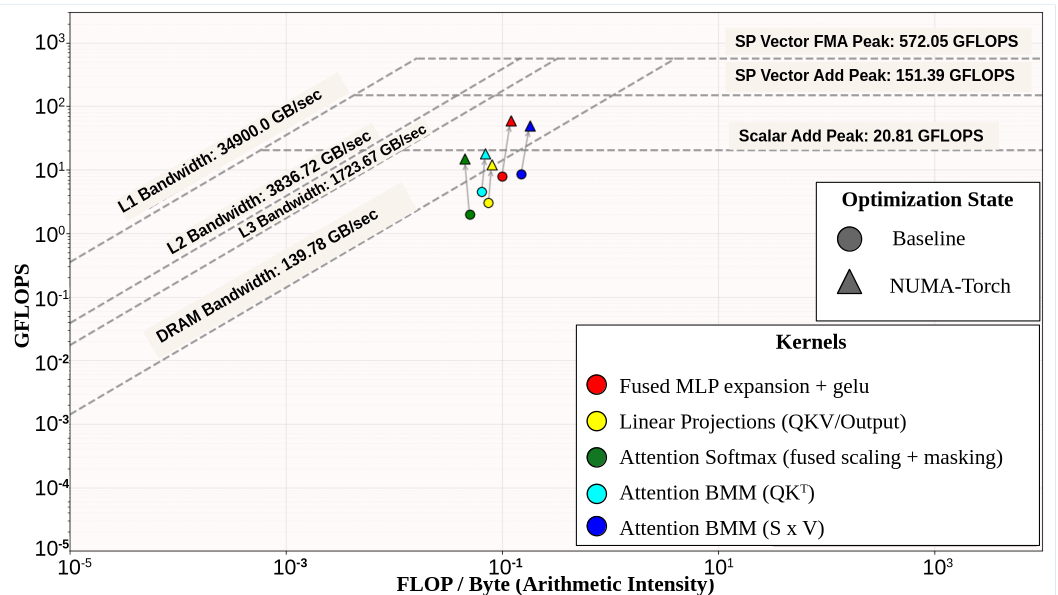}
    \caption{Roofline analysis on dual-socket Haswell: baseline Triton CPU kernels (circles) versus the \NumaTorch{} path (triangles) for the dominant GPT-2 345M prefill kernels (batch size~1, sequence length~4096). Arrows indicate the per-kernel trajectory from baseline to \NumaTorch{}.}
    \label{fig:roofline}
\end{figure}

In standard execution, compute threads are distributed
across all cores without regard to data locality, causing roughly half
of all weight accesses to incur remote DRAM latency and saturating the
inter-socket link. To empirically validate this, we conducted a Roofline
analysis using Intel Advisor~\cite{intel_advisor} on the kernels that dominate
the prefill phase of a GPT-2 345M model~\cite{radford2019gpt2} (batch size~1,
sequence length~4096) on a dual-socket Haswell platform
(Figure~\ref{fig:roofline}). For TorchInductor-compiled unmodified
Triton-CPU baseline (circles), these kernels exhibit poor data locality and sit near
the main-DRAM bandwidth ceiling, confirming that the prefill operates firmly in
the DRAM-bound regime. Our \NumaTorch{} path (triangles) improves core-to-memory
affinity and, as the trajectory arrows indicate, shifts the kernels toward the
higher-bandwidth L3 and L2 cache tiers, increasing overall throughput (GFLOPS).

Despite this architectural reality, existing CPU optimization efforts
frequently overlook the NUMA penalty, focusing on intra-core compute
optimizations. Addressing this NUMA-induced DRAM bandwidth wall must instead be
the primary objective, and the most direct solution is strict enforcement of
data locality: explicitly placing weights in socket-local DRAM via
\textit{NUMA-aware memory allocation}, and binding computations to the
corresponding cores via \textit{NUMA-aware thread scheduling}.


While mainstream frameworks for CPU inference, such as
\texttt{llama.cpp}~\cite{gerganov2023llama}, deliver high baseline
performance, they are not explicitly optimized for NUMA data locality.
Recent work such as ArcLight~\cite{xu2026arclight} addresses this
memory bottleneck through cross-NUMA tensor parallelism. However,
because ArcLight is a custom inference engine written entirely in C++,
it relies heavily on manual engineering and requires the model code to
be meticulously rewritten to support different LLM architectures. In
this work, we focus
on lifting cross-NUMA Tensor Parallelism  into PyTorch 2.0's compiler
ecosystem~\cite{ansel2024pytorch2} to provide a PyTorch-native
implementation. By extending the standard TorchInductor pipeline,
our transformations remain entirely transparent to the user, requiring
zero modifications to the model code, optimizer, or the inference loop.

We present the first compiler-based, PyTorch-native implementation of
NUMA-aware tensor parallelism targeting multi-socket CPUs. We make the
following contributions:

\begin{enumerate}
    \item A compiler-driven solution to the NUMA DRAM bottleneck that
    leverages post-grad FX graph transformations to partition
    operations across sockets, minimizing manual engineering overhead.
    \item An extended Triton-CPU backend~\cite{tillet2019triton,triton_cpu_2024} that emits per-socket buffer allocations and dispatches kernels via a persistent, NUMA-aware thread pool.
    \item An evaluation of the compiler approach across standard models (GPT-2~\cite{radford2019gpt2}, BERT~\cite{devlin2019bert}) and modern LLMs with complex attention and MLP variants (Gemma~3~\cite{team2024gemma}), showing up to $2.63\times$ prefill speedup on dual-socket platforms, comparing against process-level NUMA policies (\texttt{numactl}), and assessing scalability from dual-socket to four-socket platforms.
\end{enumerate}

\noindent This work targets the \emph{prefill} phase at fixed shapes. Throughout, ``end-to-end prefill'' denotes a single prefill forward pass at a
fixed shape, not end-to-end serving: we measure prefill in isolation, so
decode, continuous batching, KV-cache placement, and serving-level metrics are
outside the evaluation.

\begin{figure}[t]
    \centering
    \resizebox{\linewidth}{!}{
\begin{tikzpicture}[
  font=\scriptsize,
  stage/.style={draw, align=center, minimum width=2.2cm, minimum height=5.5mm,
                inner sep=2.5pt, rounded corners=1pt},
  ybox/.style={stage, fill=yellow!30, draw=orange!70!black},
  bbox/.style={stage, fill=blue!9,  draw=blue!55},
  pell/.style={ellipse, draw=red!55, fill=red!10, align=center, inner sep=1.5pt},
  gbox/.style={stage, fill=green!13, draw=green!45!black, minimum width=2.85cm},
  gboxin/.style={stage, fill=teal!22, draw=teal!60!black, minimum width=2.35cm,
                 minimum height=4.5mm},
  flow/.style={-{Triangle[width=3.2mm,length=2.2mm]}, line width=1.5mm,
               orange!90!yellow, shorten >=0.5mm, shorten <=0.5mm},
  link/.style={-{Stealth[length=1.8mm]}, semithick},
  duplex/.style={{Stealth[length=1.8mm]}-{Stealth[length=1.8mm]}, thick, red!75!black},
  msg/.style={-{Stealth[length=2mm]}, thick, orange!85!black, dashed},
  msglabel/.style={font=\scriptsize\itshape, text=orange!60!black, align=center,
                   inner sep=1pt},
]

\node[ybox] (compile) {\ttfamily @torch.compile};
\node[ybox,  below=3mm of compile] (dynamo)   {Torch Dynamo};
\node[bbox,  below=6.5mm of dynamo] (decomp)  {Decompositions\\(PrimTorch)};
\node[pell,  below=4mm of decomp]  (fx)       {\textbf{FX Passes}\\(Graph-Level Opt.)};

\coordinate (rc) at (5.0,0);
\node[gbox] (attn) at (fx -| rc) {NUMA Attention pass};
\node[gbox, above=2mm of attn] (mlp)  {NUMA MLP pass};
\node[gbox, below=2mm of attn] (norm) {NUMA Normalisation pass};

\draw[link] (fx.east) -- (mlp.west);
\draw[link] (fx.east) -- (attn.west);
\draw[link] (fx.east) -- (norm.west);

\node[align=left, inner sep=4pt, font=\tiny,
      below=6mm of norm, xshift=4mm] (schema)
  {\textbf{\scriptsize NUMA partitioning schema}\hspace{2.5mm}\mbox{}\\[0.2mm]
   {\itshape (per-block metadata)}\\[0.6mm]
   block type: MLP / Attn / Norm\\
   sockets: $N$; placement policy:\\
   \enspace 1.~column-parallel\\
   \enspace 2.~row-parallel\\
   \enspace 3.~replicated};
\newdimen\schemafold
\schemafold=2.4mm
\begin{scope}[on background layer]
\fill[orange!16]
      ($(schema.north east)+(-\schemafold,0)$) -- (schema.north west) --
      (schema.south west) -- (schema.south east) --
      ($(schema.north east)+(0,-\schemafold)$) -- cycle;
\end{scope}
\draw[orange!75!black, thick]
      ($(schema.north east)+(-\schemafold,0)$) -- (schema.north west) --
      (schema.south west) -- (schema.south east) --
      ($(schema.north east)+(0,-\schemafold)$) -- cycle;
\draw[orange!75!black, thick, fill=orange!40]
      ($(schema.north east)+(-\schemafold,0)$) --
      ($(schema.north east)+(0,-\schemafold)$) --
      ($(schema.north east)+(-\schemafold,-\schemafold)$) -- cycle;

\node[font=\scriptsize\bfseries, below=8mm of schema.south, align=center,
      anchor=north] (matitle) {NUMA-aware memory allocation};
\node[gboxin, below=1mm of matitle] (cache) {NUMA cache};
\node[gboxin, below=1.6mm of cache] (clone) {NUMA extended clone};
\begin{scope}[on background layer]
\node[draw=olive!70!black, fill=yellow!22, rounded corners=1pt, inner sep=2.5pt,
      fit=(cache)(clone)(matitle)] (memalloc) {};
\end{scope}

\draw[msg] (norm.south) -- (norm.south |- schema.north)
      node[msglabel, midway, anchor=east, xshift=-1.5mm]
      {emitted by\\NUMA passes};
\draw[msg] (norm.south |- schema.south) -- (norm.south |- memalloc.north)
      node[msglabel, midway, anchor=east, xshift=-1.5mm]
      {consumed at\\wrapper codegen};

\node[bbox] (lower) at ($(decomp |- schema.north)+(0,-1.5mm)$)
      {Lowering and scheduling\\(Inductor IR)};
\node[bbox, below=5mm of lower] (kcodegen) {Triton Kernel Codegen};
\node[pell] (wcodegen) at (decomp |- memalloc) {\textbf{Wrapper Codegen}};
\node[pell, below=13mm of wcodegen] (launcher) {\textbf{Triton CPU Launcher}};

\draw[flow] (compile) -- (dynamo);
\draw[flow] (dynamo)  -- (decomp);
\draw[flow] (decomp)  -- (fx);
\draw[flow] (fx)      -- (lower);
\draw[flow] (lower)   -- (kcodegen);
\draw[flow] (kcodegen)-- (wcodegen);
\draw[flow] (wcodegen)-- (launcher);

\draw[duplex] (wcodegen.east) -- (wcodegen.east -| memalloc.west);

\node[gbox] (sched) at (launcher -| rc) {NUMA-aware thread scheduling};
\draw[duplex] (launcher.east) -- (sched.west);

\begin{scope}[on background layer]
\node[draw=black, dashed, inner xsep=4.5mm, inner ysep=3mm,
      fit=(decomp)(fx)(lower)(kcodegen)(wcodegen)(mlp)(norm)(schema)(memalloc)]
      (inductor) {};
\node[anchor=south west, font=\scriptsize\bfseries, inner sep=1pt, fill=white]
      at ($(decomp.south |- inductor.south)+(2mm,0.6mm)$) {TorchInductor};
\node[draw=black, dashed, inner sep=2.5mm,
      fit={(launcher)(sched)($(launcher.south)+(0,-3.5mm)$)}] (runtime) {};
\node[anchor=south west, font=\scriptsize\bfseries, inner sep=1pt, fill=white]
      at ($(runtime.south west)+(0.8mm,0.6mm)$) {Triton CPU Runtime};
\end{scope}
\node[draw=red, dashed, very thick, rounded corners=2pt, inner sep=2mm,
      fit=(mlp)(norm)(schema)(memalloc)(sched)] (numatorch) {};
\node[anchor=north east, font=\scriptsize\bfseries, text=red, fill=white,
      inner sep=1.5pt] at ($(numatorch.north east)+(-1mm,-0.5mm)$) {\NumaTorch{}};

\end{tikzpicture}}
    \caption{Overview of \NumaTorch{}: \NumaTorch{} extends
    TorchInductor and Triton-CPU compilation stacks with NUMA-aware tensor
    parallel graph transformations, per-socket buffer placement, and a NUMA-aware runtime that dispatches kernels to
    socket-pinned worker groups. The NUMA partitioning schema
    (folded-corner note) is not a pipeline stage: it is per-block
    metadata emitted by the NUMA passes at FX-graph level and
    consumed by the wrapper code generator when materializing
    socket-local buffers (dashed arrows).}
    \label{fig:numa_torch_overview}
\end{figure}
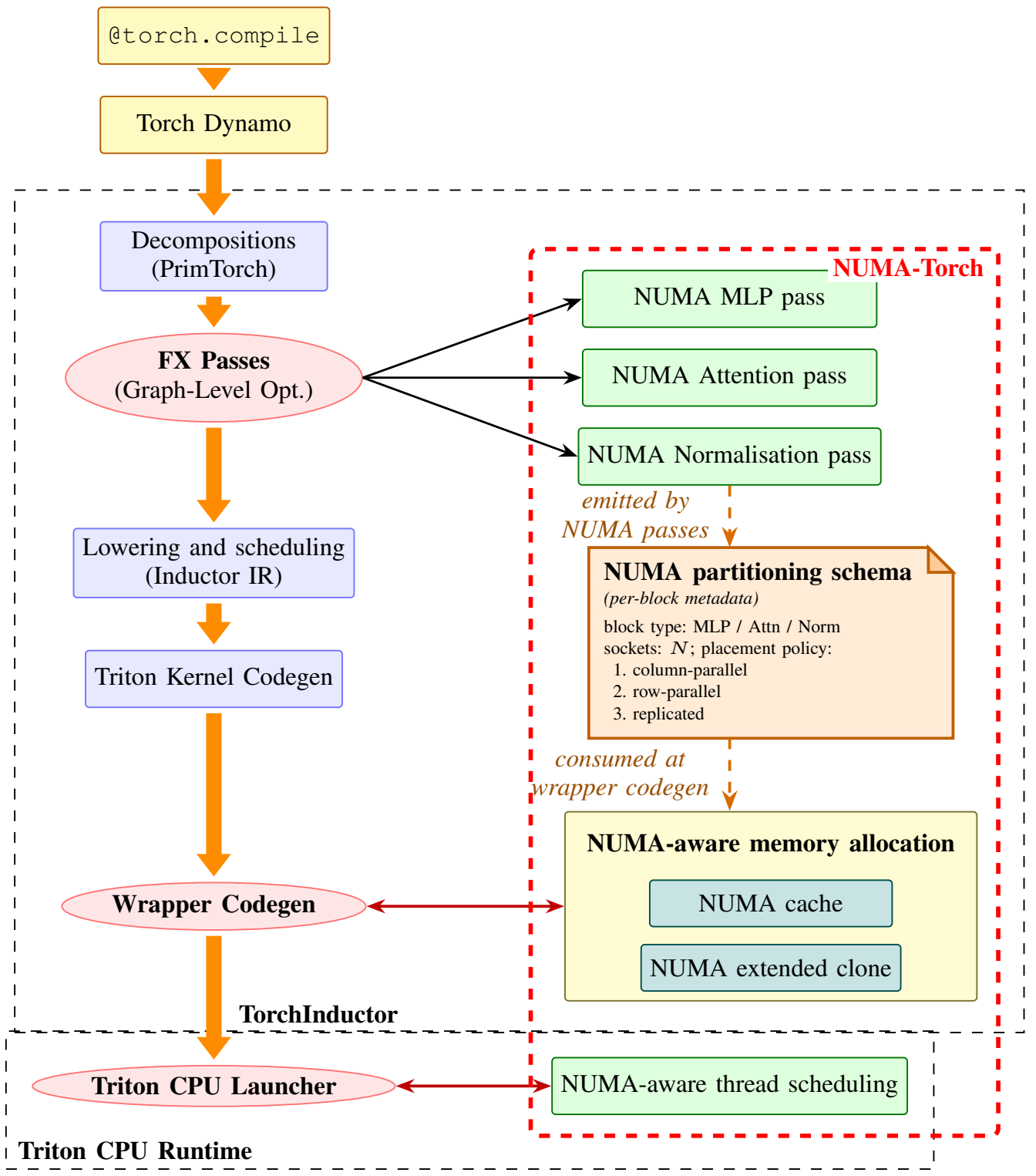

Figure~\ref{fig:numa_torch_overview} illustrates the overall design of
our approach, which we refer to as \textit{\NumaTorch{}}. \NumaTorch{} is
built on top of TorchInductor and Triton-CPU, extending the compiler
pipeline with NUMA-aware graph transformations, socket-local buffer
management, and affinity-aware kernel dispatch. Section~\ref{sec:methodology} describes each component of this pipeline in detail.

Section~\ref{sec:experiments} provides a comprehensive evaluation of \NumaTorch{}, Section~\ref{sec:related-work} discusses related literature, and we present concluding remarks in Section~\ref{sec:conclusion}. 

\section{\NumaTorch{}}\label{sec:methodology}

\subsection{Explicit NUMA Placement via Extended Clone Semantics}

Effective NUMA-aware tensor parallelism requires precise control over where a
tensor's physical memory resides. In standard PyTorch, the \texttt{clone}
operator creates a semantically independent copy. However, it does not let callers specify which socket's DRAM wiill hold the copy. Instead, cloned shards follow the calling
thread's first-touch allocation policy~\cite{linux_numa_policy_2026}, so
physical pages land in the memory bank of whichever socket's thread first
writes each virtual address. This typically concentrates storage in one
socket's DRAM, causing severe cross-interconnect traffic during subsequent
parallel computation.

To resolve this limitation, we extend the native PyTorch \texttt{clone}
operator by introducing an explicit NUMA placement parameter. This modification
empowers the runtime to bypass default allocation policies~\cite{linux_numa_policy_2026} and assign each
tensor replica to a specific DRAM socket at the exact moment of creation,
leveraging the Linux \texttt{mbind} system call~\cite{linux_mbind_2026}.

This extended functionality provides a straightforward, Pythonic interface for
tensor sharding across socket DRAMs. For example, a global weight matrix can be
easily partitioned along its rows and distributed across multiple sockets using
the new \texttt{numa\_node} argument:

\begin{lstlisting}[language=Python,basicstyle=\footnotesize\ttfamily,breaklines=true]
# Global weight matrix; defaults to first-touch allocation
weight = torch.randn(4096, 4096)

# Place tensor partitions across distinct DRAM sockets
shard_0 = weight[:2048, :].clone(numa_node=0)
shard_1 = weight[2048:, :].clone(numa_node=1)
\end{lstlisting}

\subsection{NUMA-Aware Tensor Parallelism}
\label{sec:numa_aware_scheduling}

With NUMA-aware data allocation established, the subsequent challenge is
NUMA-aware computation scheduling. Rather than relying on manual engineering,
we use the compiler to automatically identify and schedule computations, which
requires a robust underlying scheduling model. To formulate this model, a
multi-socket CPU can be abstracted as a distributed system. Each socket acts as an independent node consisting of its compute cores and local DRAM. The sockets are connected by an interconnect such as Intel’s QuickPath Interconnect (QPI) or Ultra Path Interconnect (UPI).

In distributed multi-GPU cluster environments, a fundamental goal is to reduce communication overhead to accelerate both training and inference. To mitigate these bottlenecks, works such as Megatron-LM~\cite{shoeybi2019megatron} utilize tensor parallelism to minimize synchronization points within the Multi-Layer Perceptron (MLP) and attention blocks~\cite{vaswani2017attention}. Similarly, the primary performance bottleneck in our multi-socket CPU topology is the latency and bandwidth limitation of cross-socket communication. Therefore, we adapt this tensor parallelism strategy to our single multi-socket CPU environment to minimize cross-NUMA socket communication.

We now detail how NUMA-aware Tensor Parallelism (NUMA-TP) is applied to the MLP, multi-head attention, and normalization sub-blocks that comprise a transformer layer.

\vspace{\baselineskip}
\subsubsection{MLP Block}
\label{sec:numa_tp_mlp}

Consider the MLP block of GPT-2, which consists of two
GEMMs separated by a GeLU activation, $Y = \mathrm{GeLU}(XA)$ followed by
$Z = YB$, where $X$ is the input activation and $A$, $B$ are the weight matrices.
As shown in Algorithm~\ref{alg:numa_tp_mlp}, we partition the up-projection $A$
column-wise and the down-projection $B$ row-wise. Each socket~$i$ computes its
shard of the intermediate activation $Y^i = \mathrm{GeLU}(XA^i)$ and the matching
partial output $Z^i = Y^i B^i$ entirely from socket-local memory. Because matrix
multiplication distributes over addition, the partial outputs are combined with a
single cross-socket all-reduce, $Z = \sum_i Z^i$, confining inter-socket traffic
to one activation-sized reduction per block.

\begin{algorithm}[tbh]
\footnotesize
\DontPrintSemicolon
\caption{NUMA-TP MLP Block}
\label{alg:numa_tp_mlp}
\KwIn{activation $X$; up-projection $A$; down-projection $B$; socket count $N$}
\KwOut{output activation $Z$}
Partition $A$ column-wise and $B$ row-wise across the $N$ sockets\;
\ForEach{socket $i \in \{0,\dots,N-1\}$ \textbf{\textup{in parallel}}}{
  $Y^i \leftarrow \mathrm{GeLU}(X A^i)$ \tcp*[r]{column-parallel up-proj.}
  $Z^i \leftarrow Y^i B^i$ \tcp*[r]{row-parallel down-proj.}
}
$Z \leftarrow \sum_{i=0}^{N-1} Z^i$ \tcp*[r]{cross-socket all-reduce}
\Return{$Z$}
\end{algorithm}

\subsubsection{Multi-Head Attention Block}
\label{sec:numa_tp_attention}

The same column-then-row split extends to multi-head attention, following the
Megatron-LM~\cite{shoeybi2019megatron} decomposition. The query, key, and value
projections $W_Q, W_K, W_V$ are partitioned column-wise so that each socket owns a
disjoint subset of attention heads, while the output projection $W_O$ is
partitioned row-wise. Because heads are independent within scaled dot-product
attention, every socket evaluates its local attention without exchanging queries,
keys, values, or scores across sockets. The block then closes with a single all-reduce over the per-socket partial outputs, as detailed in Algorithm~\ref{alg:numa_tp_attention}.

\begin{algorithm}[thb]
\footnotesize
\DontPrintSemicolon
\caption{NUMA-TP Multi-Head Attention Block}
\label{alg:numa_tp_attention}
\KwIn{activation $X$; projections $W_Q, W_K, W_V, W_O$; output bias $b_O$; per-head dimension $d_k$; socket count $N$}
\KwOut{output activation $Y$}
Partition $W_Q, W_K, W_V$ column-wise and $W_O$ row-wise across the $N$ sockets\;
\ForEach{socket $i \in \{0,\dots,N-1\}$ \textbf{\textup{in parallel}}}{
  $Q^i, K^i, V^i \leftarrow X W_Q^i,\; X W_K^i,\; X W_V^i$\;
  $Z^i \leftarrow \mathrm{softmax}\!\big(Q^i (K^i)^\top / \sqrt{d_k}\big)\, V^i$ \tcp*[r]{owned heads}
  $Y^i \leftarrow Z^i W_O^i$ \tcp*[r]{row-parallel out-proj.}
}
$Y \leftarrow \sum_{i=0}^{N-1} Y^i + b_O$ \tcp*[r]{all-reduce, add bias}
\Return{$Y$}
\end{algorithm}

\subsubsection{Normalization Block}
\label{sec:numa_tp_attention_layernorm}

Layer normalization and RMS normalization require a different strategy. Instead of computing the normalization on a single socket and broadcasting the result, NUMA-TP replicates this computation on every socket, as shown in Algorithm~\ref{alg:numa_tp_norm}. Each socket maintains its own normalized copy and feeds it directly to the next local kernel. This completely eliminates the need for cross-NUMA data movement, such as an all-reduce operation. We selected this strategy empirically on multi-socket platforms, where this redundant computation proved cheaper than the eliminated inter-socket communication.

\begin{algorithm}[tbh]
\footnotesize
\DontPrintSemicolon
\caption{NUMA-TP Normalization Block (LayerNorm / RMSNorm)}
\label{alg:numa_tp_norm}
\KwIn{activation $X$; normalization parameters $\gamma, \beta$; socket count $N$}
\KwOut{per-socket normalized activations $\{Y^0, \dots, Y^{N-1}\}$}
Replicate $\gamma, \beta$ to all sockets \tcp*[r]{weights not sharded}
\ForEach{socket $i \in \{0,\dots,N-1\}$ \textbf{\textup{in parallel}}}{
  $X^i \leftarrow X$ \tcp*[r]{local input copy}
  $Y^i \leftarrow \mathrm{Norm}(X^i;\, \gamma, \beta)$ \tcp*[r]{replicated full norm}
}
\Return{$\{Y^0, \dots, Y^{N-1}\}$} \tcp*[r]{no all-reduce}
\end{algorithm}

\subsection{FX Graph Transformation for Tensor Parallelism}
\label{sec:fx_graph_transformation}

Since our approach focuses on applying tensor parallelism within the PyTorch
compilation pipeline, the standard PyTorch definition of the MLP block, shown in
Figure~\ref{fig:mlp_forward}, serves as our starting point.
Here, $s$ denotes the sequence length, $h$ the input hidden dimension, $d$ the
MLP intermediate dimension, and $h_{\text{out}}$ the output dimension.

\begin{figure}[t]
\begin{lstlisting}[language=Python,basicstyle=\footnotesize\ttfamily,breaklines=true]
def forward(self, x):
    # Input activation 'x' has shape [s, h]

    # 1. Up-projection: Y = XA
    # - linear_1 applies weight A of shape [h, d]
    # - Output y expands to intermediate shape [s,d]
    y = self.linear_1(x)

    # 2. Element-wise activation
    # - Maintains the shape [s, d]
    y = torch.nn.functional.gelu(y)

    # 3. Down-projection: Z = YB
    # - linear_2 applies weight B of shape [d,h_out]
    # - Output z is projected to the terminal shape [s, h_out]
    z = self.linear_2(y)

    return z
\end{lstlisting}
\caption{Standard PyTorch \texttt{forward} definition of the MLP block.}
\label{fig:mlp_forward}
\end{figure}

\begin{figure*}[tbh]
    \centering
    \includegraphics[scale=0.35]{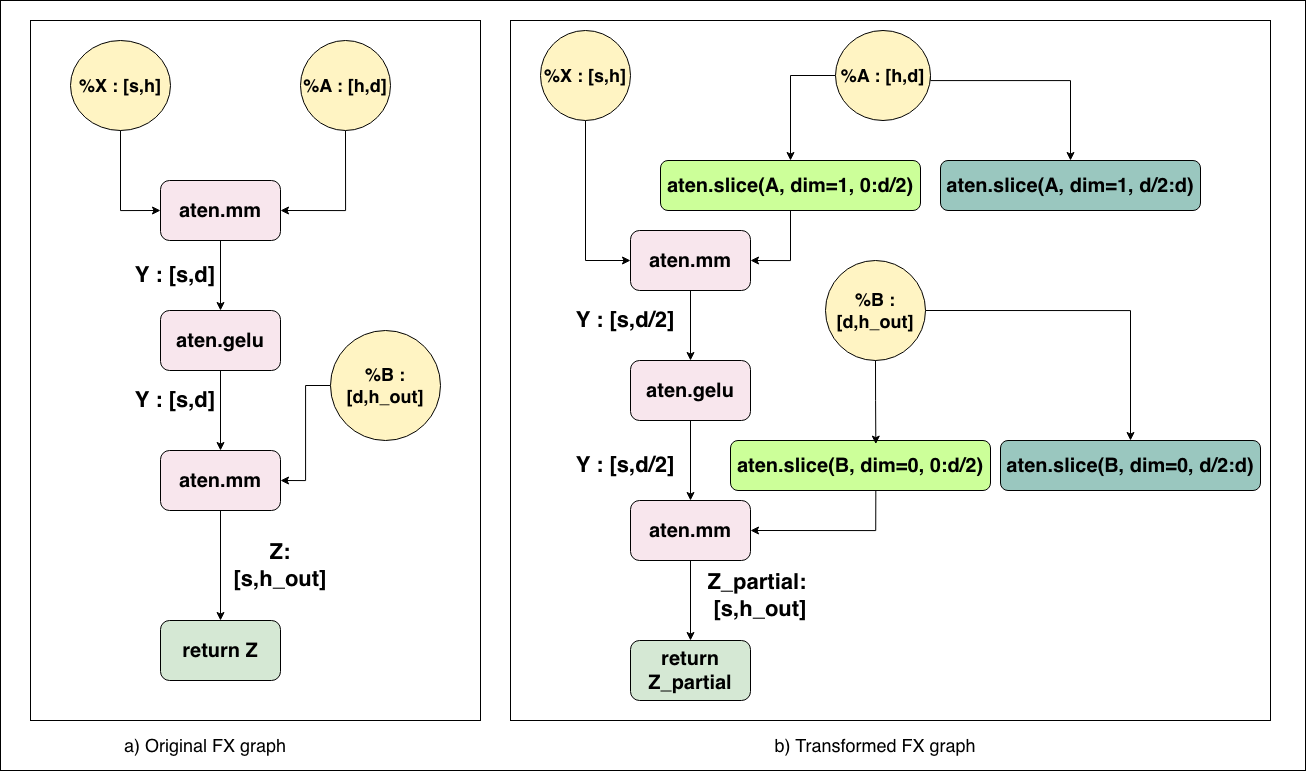}
    \caption{FX graph transformation for NUMA-aware tensor parallelism: (a) The baseline AOTAutograd FX graph for the MLP block contains an up-projection, decomposed GeLU activation, and down-projection. (b) The NUMA-TP post-grad pass rewrites the graph by inserting slice nodes that encode per-socket weight shards and shard-plan metadata, allowing physical placement and per-socket execution to be deferred to code generation.}
    \label{fig:fx_graph}
\end{figure*}

As illustrated in Figure~\ref{fig:fx_graph}a, TorchDynamo~\cite{ansel2024pytorch2} traces this module (Figure~\ref{fig:mlp_forward}) into an FX graph that sequentially projects an $[s, h]$ input to an $[s, d]$ intermediate activation, applies a decomposed pointwise GeLU subgraph, and projects it to a final $[s, h_{\text{out}}]$ output.

To apply tensor parallelism at FX graph level, we
rely on a custom post-grad FX pass that captures the canonical
\texttt{Linear} $\rightarrow$ \texttt{Activation} $\rightarrow$ \texttt{Linear}
pattern and transforms its execution topology, as illustrated in
Figure~\ref{fig:fx_graph}b. To illustrate this transformation, consider a dual-socket CPU ($N=2$) similar to the depiction in Figure~\ref{fig:numa-topology}. The parallelization pass injects \texttt{aten.slice.Tensor} nodes into the fx graph (Figure~\ref{fig:fx_graph}b). Fundamentally, \texttt{aten.slice.Tensor} is an Aten library operation~\cite{paszke2019pytorch} that extracts a contiguous sub-region of a tensor along a specified dimension. 

As discussed in Section~\ref{sec:numa_aware_scheduling}, the up-projection weight matrix $A \in \mathbb{R}^{h \times d}$ is sliced along its columns (the $d$ axis) to yield local shards of shape $[h, d/N]$. For a generalized $N$-socket system, this is achieved via the operation:
\begin{equation}
    \text{slice}\left(A, \text{dim}=1, \text{start}=i \cdot \frac{d}{N}, \text{end}=(i+1) \cdot \frac{d}{N}\right)
    \label{eq:slice_a}
\end{equation}
where $i \in \{0, \dots, N-1\}$ denotes the target NUMA socket index for that specific compute shard. 

Similarly, for the down-projection, the weight matrix $B \in \mathbb{R}^{d \times h_{\text{out}}}$ is sliced along its rows (the $d$ axis) to produce shards of shape $[d/N, h_{\text{out}}]$. The corresponding operation is:
\begin{equation}
    \text{slice}\left(B, \text{dim}=0, \text{start}=i \cdot \frac{d}{N}, \text{end}=(i+1) \cdot \frac{d}{N}\right)
    \label{eq:slice_b}
\end{equation}

As established in Section~\ref{sec:numa_aware_scheduling}, yielding the correct final output $Z$ requires an element-wise reduction (addition) of the partial results computed by each socket. However, the $N$-way reduction node is absent from the transformed FX graph in Figure~\ref{fig:fx_graph}b. 

This omission is a deliberate architectural choice, rooted in the fact that the FX graph maintains a ``Single-Program, Multiple-Data'' (SPMD) representation. The graph does not instantiate $N$ separate matrix multiplication (\texttt{mm}) nodes; instead, there is exactly one logical \texttt{mm} node whose dimension metadata has been scaled down from the global $[s, d]$ to the local $[s, d/N]$. Because the graph encodes the execution logic for a single generic shard, there are no $N$ distinct partial-sum nodes in the intermediate representation to combine.

Consequently, the physical parallelization and the $N$-way reduction are deferred entirely to the code-generation phase (discussed in Section~\ref{sec:d}). The actual existence of $N$ separate partial outputs ($Z^{0} \dots Z^{N-1}$) only manifests at runtime when the wrapper codegen allocates $N$ independent buffers and launches the identical kernel $N$ times across the hardware domains.

\subsection{The NUMA Partitioning Schema: Bridging Graph Transformations and Code Generation}
\label{sec:numa_partitioning_schema}

The NUMA partitioning schema is the metadata bridge between the mathematical rewriting performed in the FX graph and the physical multi-socket allocation performed by the backend code generator. As depicted in Figure~\ref{fig:numa_torch_overview}, it records for each tensor its placement policy (\emph{column-parallel}, \emph{row-parallel}, or \emph{replicated}), which the wrapper code generator later uses to materialize socket-local buffers. In the MLP block above, the up-projection $A$ is \emph{column-parallel}, the down-projection $B$ \emph{row-parallel}, and the input activation $X$ \emph{replicated} across all sockets.

\subsection{Wrapper Code Generation and Buffer Materialization}
\label{sec:d}


So far the partitioning exists only as metadata, with no buffer allocated and no page bound to a socket. Both happen in the code TorchInductor generates, which is Triton-CPU kernel code~\cite{triton_cpu_2024} together with the Python wrapper code that drives it. The wrapper function call allocates each buffer, places it, and only then launches the kernels that read it. For every kernel argument it consults the schema, classifies the tensor as a column shard, a row shard, or a full replica, and emits the matching slice with a \texttt{clone(numa\_node=i)}. That explicit \texttt{numa\_node} binds shard~$i$'s pages to socket~$i$'s DRAM at creation instead of leaving them to first touch. Because each placement finishes before the kernel that reads it is launched, every socket's kernel finds its operands already local. The FX pass therefore states only how a block is partitioned; wrapper code generation is where its shards first acquire physical pages.

This stage also decides which socket-local buffers can be reused across adjacent
kernels. When an activation produced on socket $i$ is consumed immediately by
another NUMA-parallel kernel on the same socket, the wrapper records it in a
per-socket cache instead of constructing a global tensor and slicing it again. In the MLP example, the column-parallel up-projection output
$Y^i=\mathrm{GeLU}(XA^i)$ stays local and feeds the row-parallel
down-projection consuming $B^i$, so the only cross-socket step is the
reduction that closes the block.

Because the final reduction is not explicit in the FX graph
(Figure~\ref{fig:fx_graph}b), the wrapper must know whether a kernel produces a
still-local intermediate or a value closing the NUMA-parallel region. The distinction matters most after fusion, where one lowered kernel may contain both matrix multiplication and pointwise operations. We therefore attach
an output tag to each generated value, and the wrapper uses it to choose one of
two reconciliation paths, shown for a dual-socket system in
Figure~\ref{fig:reconciliation}:

\begin{itemize}
    \item \textbf{Intermediate activations} (Figure~\ref{fig:reconciliation}a)\textbf{:} Outputs from column-parallel projections and fused pointwise operations remain socket-local. The wrapper stores each shard in the per-socket cache and forwards it to the next local kernel without reduction.
    \item \textbf{Terminal partial sums} (Figure~\ref{fig:reconciliation}b)\textbf{:} Outputs from row-parallel projections, such as the MLP down-projection or the attention output projection, represent partial values $Z^{i}$. Before such a value leaves the NUMA-parallel block, the wrapper inserts the all-reduce loop that sums the per-socket partial outputs into the final tensor $Z$.
\end{itemize}

\subsection{NUMA-Aware Kernel Dispatch and Deferred Reassembly}
\label{sec:numa_kernel_dispatch}

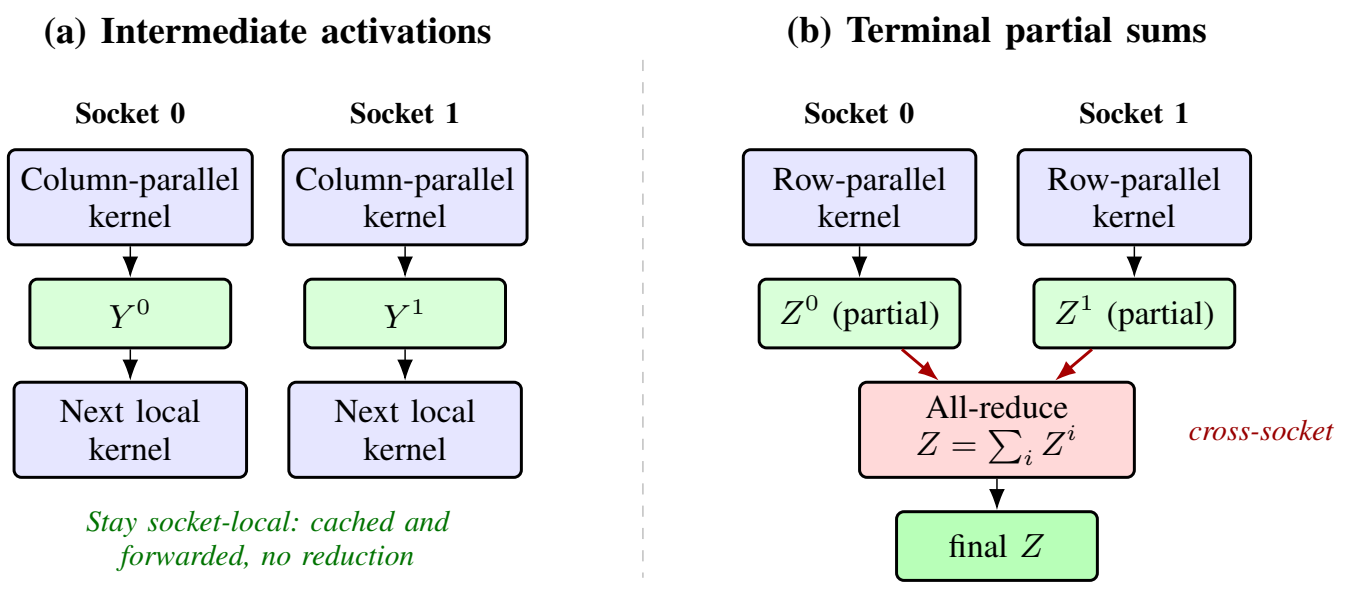
\begin{figure}[tbh]
\centering
\resizebox{\linewidth}{!}{%
\begin{tikzpicture}[
  font=\small,
  >={Latex[length=2mm]},
  hdr/.style={font=\footnotesize\bfseries},
  kern/.style={draw,thick,fill=blue!10,rounded corners=2pt,minimum width=2.2cm,minimum height=0.9cm,align=center},
  buf/.style={draw,thick,fill=green!15,rounded corners=2pt,minimum width=1.9cm,minimum height=0.65cm,align=center},
  redn/.style={draw,thick,fill=red!15,rounded corners=2pt,minimum width=2.6cm,minimum height=0.9cm,align=center},
  fin/.style={draw,thick,fill=green!28,rounded corners=2pt,minimum width=1.9cm,minimum height=0.65cm,align=center},
  note/.style={font=\footnotesize\itshape,align=center}
]

\node[hdr] at (0,2.0) {Socket 0};
\node[hdr] at (2.6,2.0) {Socket 1};
\node[kern] (ak0) at (0,1.2) {Column-parallel\\kernel};
\node[kern] (ak1) at (2.6,1.2) {Column-parallel\\kernel};
\node[buf]  (ay0) at (0,0.1) {$Y^{0}$};
\node[buf]  (ay1) at (2.6,0.1) {$Y^{1}$};
\node[kern] (an0) at (0,-1.0) {Next local\\kernel};
\node[kern] (an1) at (2.6,-1.0) {Next local\\kernel};
\draw[->] (ak0)--(ay0); \draw[->] (ay0)--(an0);
\draw[->] (ak1)--(ay1); \draw[->] (ay1)--(an1);
\node[note,text=green!45!black] at (1.3,-2.05) {Stay socket-local: cached and\\forwarded, no reduction};
\node[font=\bfseries] at (1.3,2.75) {(a) Intermediate activations};

\node[hdr] at (6.9,2.0) {Socket 0};
\node[hdr] at (9.5,2.0) {Socket 1};
\node[kern] (bk0) at (6.9,1.2) {Row-parallel\\kernel};
\node[kern] (bk1) at (9.5,1.2) {Row-parallel\\kernel};
\node[buf]  (bz0) at (6.9,0.1) {$Z^{0}$ (partial)};
\node[buf]  (bz1) at (9.5,0.1) {$Z^{1}$ (partial)};
\node[redn] (bar) at (8.2,-1.0) {All-reduce\\$Z=\sum_{i} Z^{i}$};
\node[fin]  (bz)  at (8.2,-2.1) {final $Z$};
\draw[->] (bk0)--(bz0);
\draw[->] (bk1)--(bz1);
\draw[->,red!65!black,thick] (bz0) -- (bar);
\draw[->,red!65!black,thick] (bz1) -- (bar);
\draw[->] (bar)--(bz);
\node[note,text=red!60!black] at (10.7,-1.0) {cross-socket};
\node[font=\bfseries] at (8.2,2.75) {(b) Terminal partial sums};

\draw[gray!50,dashed] (4.85,2.5) -- (4.85,-2.4);

\end{tikzpicture}%
}
\caption{The two buffer-reconciliation patterns chosen by the wrapper from
each value's semantic output tag. (a) Outputs of
column-parallel kernels and fused pointwise ops are tagged as intermediate
activations: each per-socket shard $Y^{i}$ stays in socket-local DRAM, is held in
the per-socket cache, and is forwarded to the next local kernel with no
cross-socket communication. (b) Outputs of row-parallel kernels are tagged as
terminal partial sums $Z^{i}$; before the value leaves the NUMA-parallel block,
the wrapper emits a single cross-socket all-reduce that sums the partials into the
final tensor $Z$.}
\label{fig:reconciliation}
\end{figure}

After the wrapper has placed each tensor shard in its socket-local memory,
execution is handed to the Triton-CPU launcher. In the unmodified backend, this
launcher runs a kernel as a single program replicated across many parallel work
items, but it treats the machine as one flat pool of cores with no notion of NUMA
locality. As a result, even when a shard lives in a particular socket's DRAM,
the default launcher gives no guarantee that the threads using that shard
run on cores of the same socket. Many instead run on a remote socket,
paying the interconnect penalty and undoing the placement the wrapper
performed.

We therefore extend this launch path with a NUMA-aware dispatch layer,
summarized in Algorithm~\ref{alg:numa_launcher}. Each time a kernel is launched,
the layer checks whether the compiler has marked it for NUMA execution (i.e.,
whether the wrapper has attached the per-socket operands together with a note of
which socket each shard belongs to). If this annotation is absent, the kernel is
launched through the original NUMA-oblivious path, and hence, ordinary non-sharded
operations are left untouched. If it is present, the launcher takes the
NUMA-aware path described next.

On this path, the launcher turns the single-program, multiple-data view of the FX
graph into execution along two levels of parallelism. Across sockets,
every socket runs its own copy of the kernel over the full set of work items,
reading only that socket's local shards, so no socket reaches across the
interconnect for data. Within a socket, the work is shared among a group of threads. The
launcher first fixes a total thread budget, taking either the parallelism
the kernel requests or the available hardware threads, but never fewer
than one thread per socket. It then splits that budget into $N$ roughly
equal groups, one for each socket, and binds every group's threads to
cores of its own socket as reported by the OS topology. No thread
therefore drifts onto another socket.
The launcher waits for every group to finish, so the per-socket
kernels complete together. With this binding in place, each thread runs on the socket that
physically holds the shard it touches. This is what closes the locality
gap left by the default launcher.

\begin{algorithm}[t]
\footnotesize
\DontPrintSemicolon
\caption{NUMA-Aware Kernel Launch}
\label{alg:numa_launcher}
\KwIn{kernel $K$; its set of parallel work items $G$; when NUMA-tagged, the per-socket local operands $\{P_0,\dots,P_{N-1}\}$ and the socket each shard belongs to; thread budget $T$}
\KwOut{$K$ executed, with every socket writing its output shard into local memory}
\eIf{$K$ is tagged for NUMA execution}{
  $N \leftarrow$ number of sockets in the per-socket operands\;
  $T \leftarrow \max(T, N)$\;
  Split the $T$ threads into $N$ groups of $T_i \approx T/N$, one per socket\;
  \ForEach{socket $i \in \{0,\dots,N-1\}$ \textbf{\textup{in parallel}}}{
    $C_i \leftarrow$ cores belonging to socket $i$\;
    \ForEach{worker $j$ in socket $i$'s group of $T_i$ threads}{
      bind worker $j$ to a core in $C_i$\;
      run $K$ on worker $j$'s $1/T_i$ share of $G$, reading $P_i$\;
    }
  }
  wait for all groups to finish\;
}{
  run $K$ through the default, NUMA-oblivious launch\;
}
\end{algorithm}

This dispatch policy completes the path from graph rewriting to execution: the FX
pass marks an operation as shardable, the schema records how each of its tensors
should be placed, the wrapper prepares the socket-local operands, and the launcher
runs one kernel instance per socket. Reassembly is left to the wrapper rather than
the launcher and is deferred until a value actually needs to be combined
(Figure~\ref{fig:reconciliation}). A value consumed only by another NUMA-aware
kernel stays split, so the wrapper keeps the per-socket buffers and forwards them
as they are. A value that is a final partial sum, or that is about to leave the
parallel region, is closed with the cross-socket all-reduce described earlier,
after which the reassembled tensor re-enters the ordinary PyTorch program.

\section{Experimental Evaluation}
\label{sec:experiments}

\subsection{Experimental Setup}

We evaluate \emph{\NumaTorch{}}, on three Intel Xeon platforms: two
dual-socket machines (CascadeLake and Haswell) and one four-socket machine
(Haswell-EX), the last sharing the Haswell microarchitecture with the
dual-socket Haswell server so that comparing the two isolates the number of
NUMA domains. Table~\ref{tab:hardware} summarizes the hardware and
Table~\ref{tab:software} the software stack.

\begin{table}[tbh]
\centering
\caption{Hardware specifications of the three evaluation platforms}
\label{tab:hardware}
\resizebox{\columnwidth}{!}{%
\begin{tabular}{@{}llll@{}}
\toprule
\textbf{Specification} & \textbf{CascadeLake} & \textbf{Haswell} & \textbf{Haswell-EX} \\ \midrule
Processor & Xeon Gold 6253CL & Xeon E5-2620 v3 & Xeon E7-4820 v3 \\
Microarchitecture & Cascade Lake & Haswell & Haswell \\
Total vCPUs & 60 & 24 & 80 \\
Cores per socket & 15 & 6 & 10 \\
Sockets / NUMA nodes & 2 & 2 & 4 \\
Clock speed & 3.10\,GHz & 2.40\,GHz & 1.90\,GHz \\
L3 cache & 49.5\,MiB & 30\,MiB & 100\,MiB \\
Environment & GCP \texttt{c2-standard-60} & Server (bare metal) & Server (bare metal) \\
\bottomrule
\end{tabular}%
}
\end{table}

\begin{table}[tbh]
\centering
\caption{Software configuration}
\label{tab:software}
\resizebox{\columnwidth}{!}{%
\begin{tabular}{@{}lll@{}}
\toprule
\textbf{Component} & \textbf{Dual-socket} & \textbf{Four-socket} \\ \midrule
Operating system & Ubuntu 20.04 LTS & Ubuntu 24.04 LTS \\
Linux kernel & 5.15 & 7.0 \\
Python & 3.8 & 3.9 \\
PyTorch & \multicolumn{2}{l}{2.6.0a0 (\texttt{torch.compile} / TorchInductor)} \\
Triton-CPU & \multicolumn{2}{l}{commit \texttt{7e394470}} \\
LLVM (Triton-CPU backend) & \multicolumn{2}{l}{19 (commit \texttt{adba14a})} \\
\bottomrule
\end{tabular}%
}
\end{table}

\subsection{Models and Configurations}

We evaluate three model families, summarized in
Table~\ref{tab:model_configs}.  GPT-2~\cite{radford2019gpt2}
(decoder-only) and BERT~\cite{devlin2019bert} (encoder-only) exercise the
core column- and row-parallel FX passes with standard multi-head
attention, while Gemma~3 270M~\cite{team2024gemma} additionally tests the
KV-replication path for Multi-Query Attention
(MQA)~\cite{shazeer2019mqa} ($n_{kv}\!=\!1$) and the GeGLU-gated MLP
pass.

\begin{table}[tbh]
\centering
\caption{Architectural parameters of evaluated models}
\label{tab:model_configs}
\resizebox{\columnwidth}{!}{%
\begin{tabular}{@{}llcccccc@{}}
\toprule
\textbf{Family} & \textbf{Model} & \textbf{Params} & \textbf{Layers} &
$\mathbf{d_{\rm model}}$ & \textbf{Heads ($n_h$)} &
$\mathbf{n_{kv}}$ & \textbf{MLP type} \\ \midrule
\multirow{3}{*}{GPT-2~\cite{radford2019gpt2}}
  & 117M  & 117M & 12 & 768  & 12 & 12 & GeLU \\
  & 345M  & 345M & 24 & 1024 & 16 & 16 & GeLU \\
  & 762M  & 762M & 36 & 1280 & 20 & 20 & GeLU \\
\midrule
\multirow{2}{*}{BERT~\cite{devlin2019bert}}
  & Small & 29M  & 4  & 512  & 8  & 8  & GeLU \\
  & Large & 340M & 24 & 1024 & 16 & 16 & GeLU \\
\midrule
Gemma 3~\cite{team2024gemma} & 270M & 270M & 18 & 640 & 4 & 1 & GeGLU \\
\bottomrule
\end{tabular}%
}
\end{table}

For each model we measure the mean prefill latency (ms) across a
representative sweep of batch-size and sequence-length configurations.
Each latency data point discards 10 warm-up iterations and averages the
wall-clock time of 50 timed iterations, ensuring that compilation
artifacts and cold-cache effects are excluded from reported figures.

On CPU, TorchInductor generates C++/OpenMP code by
default~\cite{ansel2024pytorch2}.  We instead select
Triton-CPU~\cite{triton_cpu_2024} as the CPU
code-generation backend through
\texttt{torch.\_inductor.config.cpu\_backend}, because NUMA-TP needs
control of the kernel launch path in order to bind thread groups to
sockets.  Both configurations below use it, so reported speedups are
relative to Triton-CPU and do not transfer as absolute latencies to the
default C++/OpenMP backend.

The baseline is \texttt{torch.compile} with unmodified TorchInductor and
the unmodified Triton-CPU backend.  \NumaTorch{} modifies both, as
described in Section~\ref{sec:methodology}.



\subsection{Performance Results}

\begin{figure}[!t]
\centering
\subfloat[Haswell (Intel Xeon E5-2620 v3, 6 cores/socket)]{%
  \includegraphics[width=0.78\columnwidth]{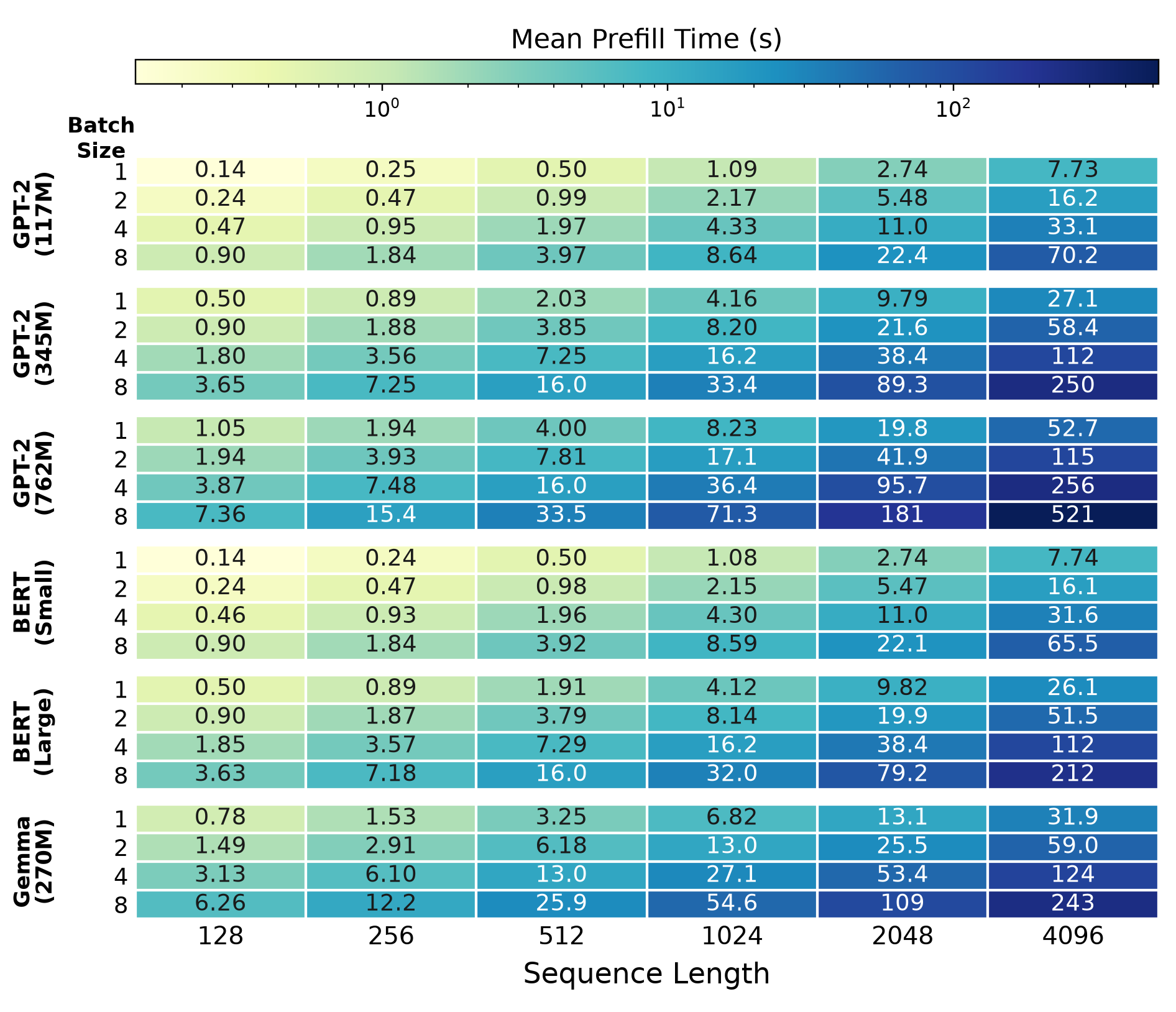}}\\
\subfloat[Cascade Lake (Intel Xeon Gold 6253CL, 15 cores/socket)]{%
  \includegraphics[width=0.78\columnwidth]{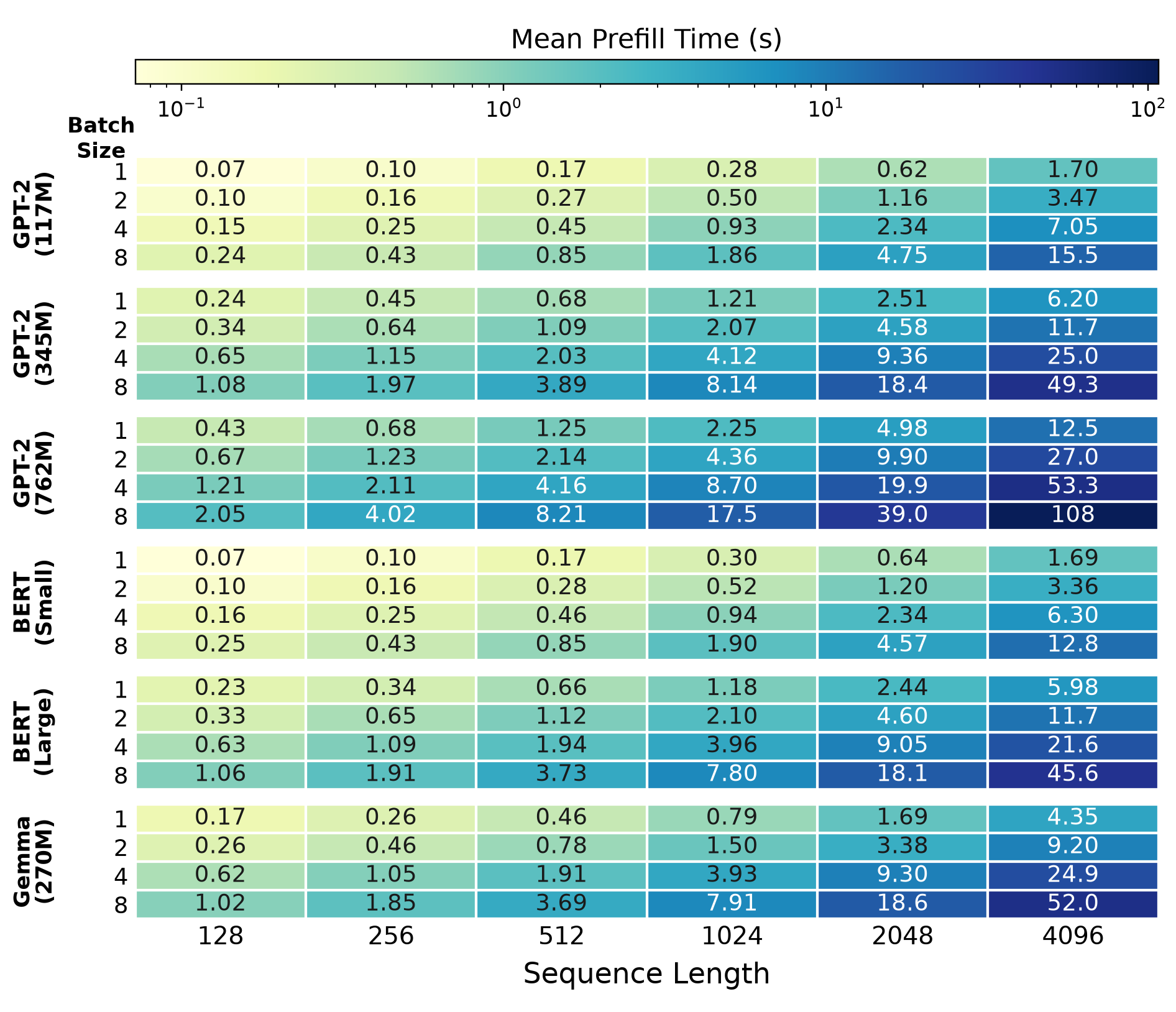}}
\caption{Mean prefill execution time of \NumaTorch{}, for every evaluated
model across the full sweep of batch sizes (rows) and sequence lengths
(columns), on the dual-socket (a)~Haswell and (b)~Cascade Lake platforms.
Each cell is annotated with the mean wall-clock time of one prefill call
in seconds; the logarithmic color scale, shown above each heatmap, is set
independently per platform. Over all measured configurations, the coefficient
of variation of prefill time has mean $1.58\%$ (max $6.74\%$) on Haswell and
mean $4.43\%$ (max $26.23\%$) on Cascade Lake.}
\label{fig:numa_time_heatmaps}
\end{figure}

The mean prefill execution times of \NumaTorch{} are presented as
per-configuration heatmaps in
Figure~\ref{fig:numa_time_heatmaps}, spanning every evaluated model across
the full sweep of batch sizes and sequence lengths on both the Haswell and
Cascade Lake architectures.

We evaluate the end-to-end prefill speedup of the \NumaTorch{} compiler
extension relative to the baseline. As established by our
initial roofline analysis, the baseline prefill phase is
bottlenecked by main memory bandwidth. By strictly enforcing data
locality via our NUMA-aware tensor parallelism passes, \NumaTorch{}
successfully mitigates cross-socket interconnect traffic, unlocking
speedups of up to $2.63\times$ on Haswell and $1.87\times$ on Cascade
Lake for all evaluated transformer architectures
(Figure~\ref{fig:speedup_bars}).

\subsubsection*{1) Architectural Impact (Haswell vs. Cascade Lake)}
The gain tracks the number of cores per socket. On Haswell (Figure~\ref{fig:speedup_bars}a), with just 6 cores per socket, speedups frequently exceed $2.0\times$ at sequence lengths of 1024 and above: with few cores, NUMA-oblivious scheduling drives a high volume of remote traffic through each core, so threads stall on QPI fetches and the baseline is highly inefficient. Cascade Lake (Figure~\ref{fig:speedup_bars}b), with 15 cores per socket, spreads that traffic across more threads, dampening the baseline's cross-socket bottleneck; speedups are accordingly more moderate but highly consistent, ranging from $1.10\times$ to $1.87\times$.

\subsubsection*{2) Impact of Model Scale and Architecture}
Larger hidden dimensions and weight matrices yield higher relative speedups, placing a heavier burden on the memory sub-system: on Haswell at batch size 8, GPT-2~\cite{radford2019gpt2} (345M) peaks at $2.63\times$ (sequence length 256) against $2.10\times$ for the smaller BERT~\cite{devlin2019bert} (Small) under identical conditions. Gemma 3~\cite{team2024gemma} (270M) combines Multi-Query Attention with
a GeGLU-gated MLP.  It improves robustly as well, reaching $1.96\times$
on Haswell and $1.76\times$ on Cascade Lake, which confirms that the FX
transformations generalize to heterogeneous sub-layers without model code
rewrites.

\subsubsection*{3) Batch Size and Sequence Length Scaling}
The optimizations scale efficiently as workload size grows. For long-context workloads intermediate activations dominate the prefill footprint.

\begin{figure*}[t]
\centering
\subfloat[Haswell (Intel Xeon E5-2620 v3, 6 cores/socket)]{%
  \includegraphics[width=\textwidth]{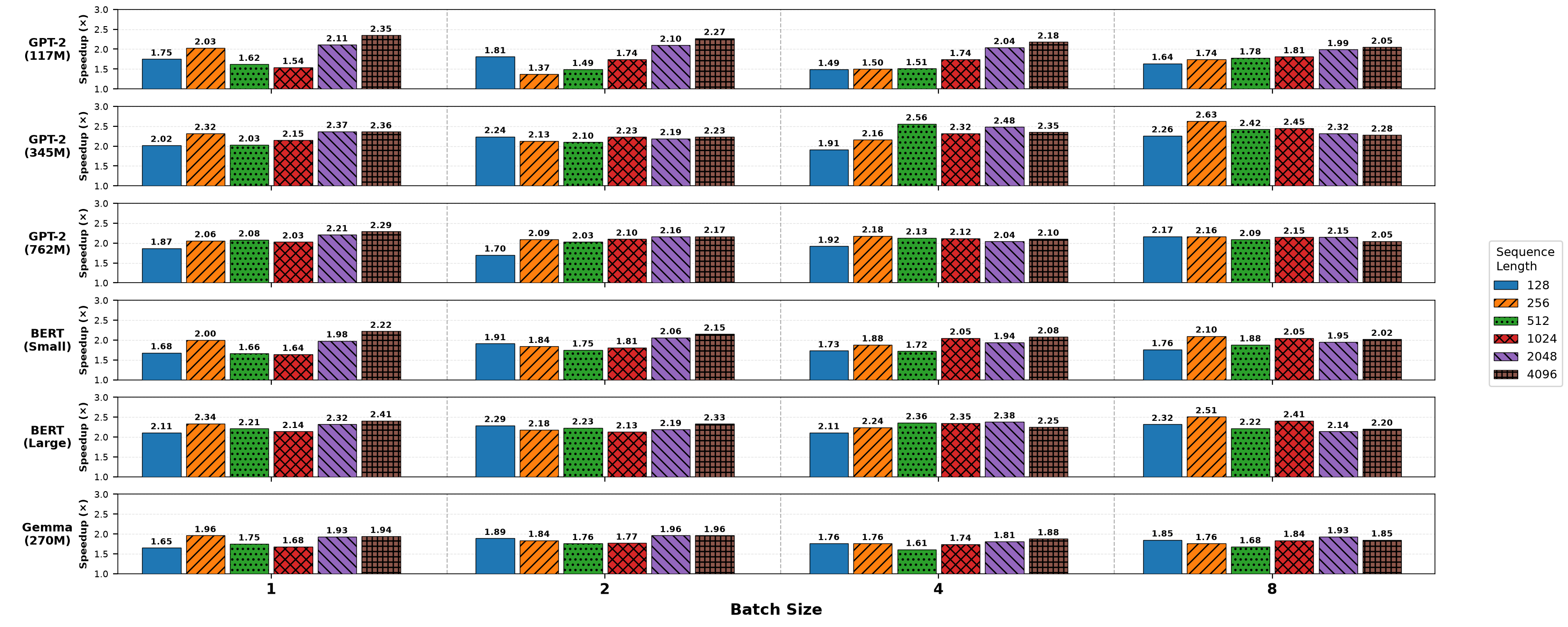}}\\
\subfloat[Cascade Lake (Intel Xeon Gold 6253CL, 15 cores/socket)]{%
  \includegraphics[width=\textwidth]{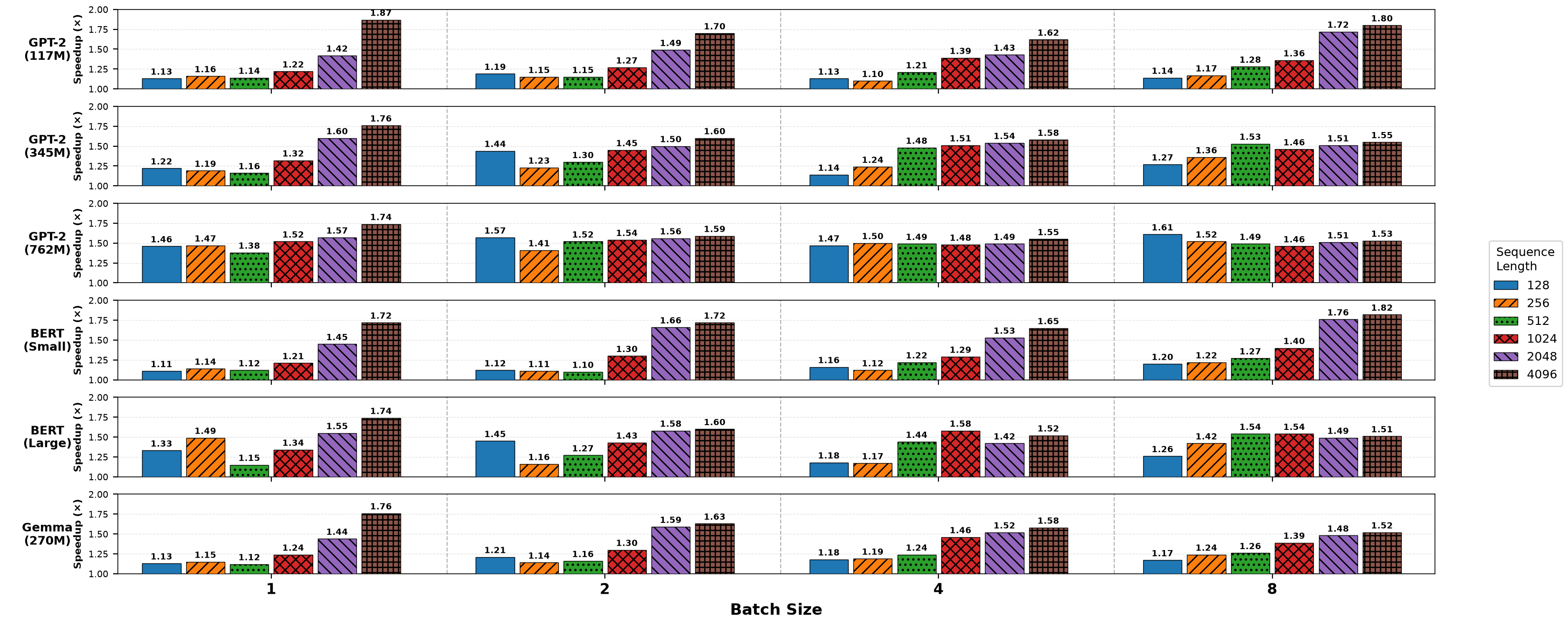}}
\caption{End-to-end prefill speedup of \NumaTorch{} over the unmodified
Triton-CPU baseline for every evaluated configuration, grouped by
model, batch size, and sequence length, on the dual-socket (a)~Haswell
and (b)~Cascade Lake platforms.}
\label{fig:speedup_bars}
\end{figure*}

Figures~\ref{fig:numa_time_heatmaps} and~\ref{fig:speedup_bars} together
report every evaluated configuration in absolute and relative terms,
where the savings at the longer sequence lengths reach multiple seconds
per prefill call.

\subsection{Comparison with Process-Level NUMA Policies}
\label{sec:numactl}

Do the operating system's own placement knobs already recover this
benefit?  We re-ran the full 144-configuration Cascade Lake sweep under
three \texttt{numactl} policies~\cite{numactl_2026}, each measured
against default first-touch allocation (Table~\ref{tab:numactl}).

None of the three comes close, and each falls short for a different
reason. Interleaving reaches a geometric-mean speedup of only
$1.08\times$ and a median speedup of $1.02\times$.  Round-robin placement equalizes which memory
controller serves a request, but no access becomes local.  It removes
bandwidth imbalance and leaves the interconnect volume that
Figure~\ref{fig:roofline} identifies as binding.  It is also the least
reliable policy, running slower than default in 44 of 144 configurations.

\texttt{localalloc} is the best of the three by geometric-mean speedup,
at $1.12\times$, and still recovers little.  Two limits explain why.  No process-level policy can
shard a tensor, so each weight matrix lands whole on one socket and the
other socket reads all of it remotely.  First-touch locality also fails
to survive execution, because TorchInductor's NUMA-oblivious thread pool
means the thread that first touches a buffer is rarely the one that
reads it.  NUMA-Torch removes both limits, pairing
\texttt{clone(numa\_node=$i$)} placement with thread-group binding.

Single-socket execution is the most informative control.  Its locality
is perfect by construction, yet its geometric-mean speedup is only
$0.59\times$, making it the worst option by a wide margin; it runs slower
than default in 143 of 144 configurations.  It
trades all remote traffic for half the cores and half the bandwidth.

NUMA-Torch reaches a geometric-mean speedup of $1.38\times$ and a median
speedup of $1.44\times$, and it is alone in never running slower than
default, doing so in 0 of 144 configurations.  The goal is not locality alone,
but locality without sacrificing the machine's full compute and memory
bandwidth.

\begin{table}[bh]
\centering
\caption{Process-level NUMA policies on Cascade Lake}
\label{tab:numactl}
\resizebox{\columnwidth}{!}{%
\begin{tabular}{@{}lccc@{}}
\toprule
\textbf{Policy} & \textbf{Geo.\ mean} & \textbf{Median} & \textbf{Worse than default} \\ \midrule
\texttt{--interleave=all} & 1.08$\times$ & 1.02$\times$ & 44/144 \\
\texttt{--localalloc}     & 1.12$\times$ & 1.06$\times$ & 19/144 \\
Single socket (\texttt{--cpunodebind=0 --membind=0}) & 0.59$\times$ & 0.59$\times$ & 143/144 \\ \midrule
NUMA-TP (this work) & \textbf{1.38}$\times$ & \textbf{1.44}$\times$ & \textbf{0/144} \\
\bottomrule
\end{tabular}%
}
\end{table}

\subsection{Scaling to Four NUMA Domains}
\label{sec:foursocket}

\begin{figure}[t]
\centering
\includegraphics[width=0.85\columnwidth]{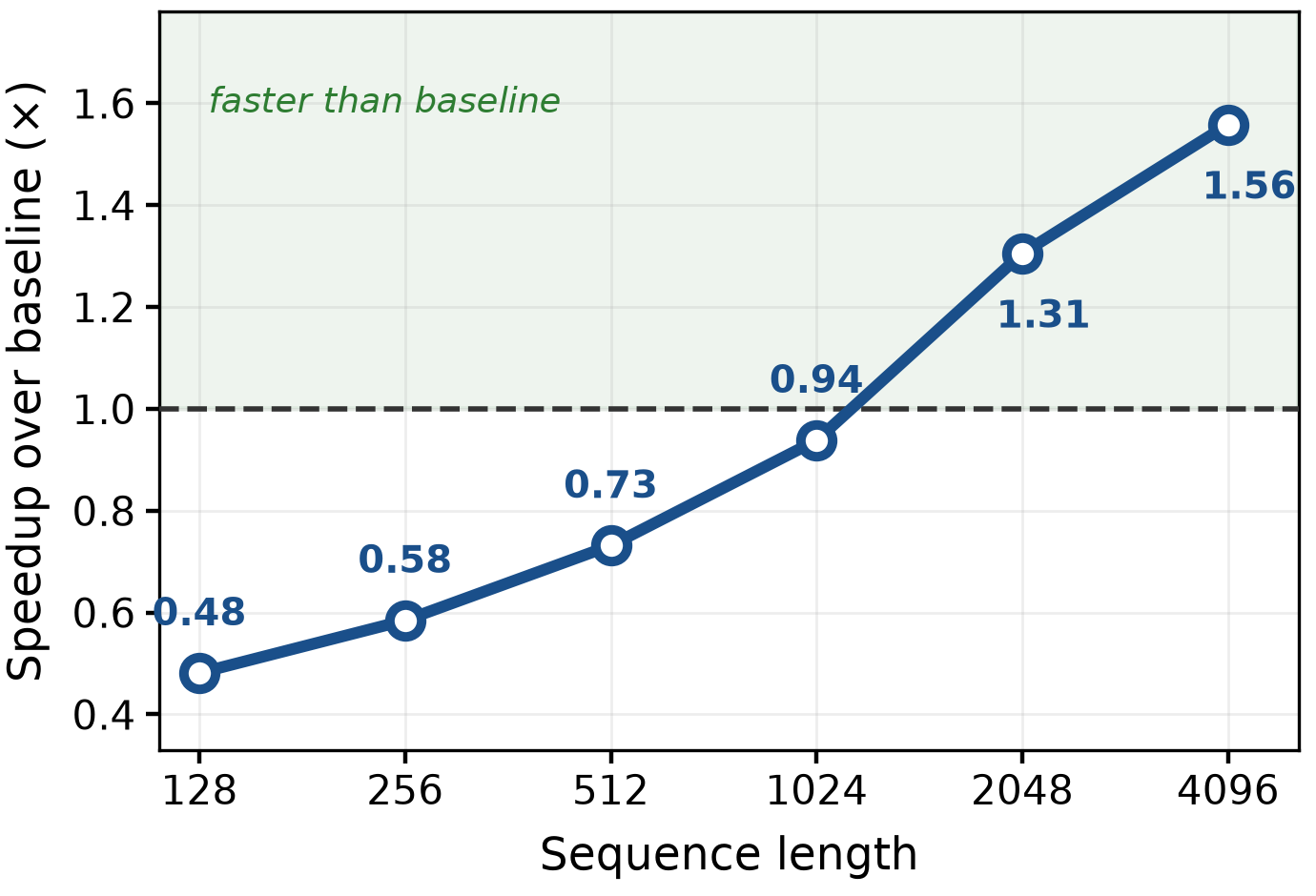}
\caption{End-to-end prefill speedup of \NumaTorch{} over the unmodified
 Triton-CPU baseline on the four-socket Haswell-EX platform, as the geometric
 mean over all models and batch sizes at each sequence length. The
dashed line marks parity; in the shaded region the geometric mean exceeds
$1.0\times$.}
\label{fig:foursocket}
\end{figure}

We repeat the evaluation on the four-socket Haswell-EX platform of
Table~\ref{tab:hardware}, sweeping the same six models over same batch size and sequence length configurations.
Figure~\ref{fig:foursocket} reports the resulting speedups.

Speedup rises monotonically with sequence length, from a geometric mean of
$0.48\times$ at $s{=}128$ to $1.56\times$ at $s{=}4096$, crossing $1.0\times$
between $s{=}1024$ and $s{=}2048$. Every measured configuration at
$s \geq 2048$ is faster than the baseline, with a geometric mean of
$1.38\times$ and a peak of $2.27\times$ for BERT~Small at $B{=}2$ and
$s{=}4096$, where the mean prefill time falls from $25.8$\,s to $11.3$\,s.
No configuration at $s \leq 512$ is faster than the baseline.
Lower speedups at the shorter sequence lengths are expected on this platform,
for two compounding reasons. Both reasons are already present on the dual-socket platforms and simply
combine unfavourably here.  The first is amortization.  Thread-group
binding, the barrier (Algorithm~\ref{alg:numa_launcher}), and the
all-reduce cost about the same per kernel whatever the kernel computes,
while the work each socket does shrinks as domains are added.  A socket
here takes a quarter of the partitioned intermediate width rather than a
half, and every reassembly point closes with a four-way all-reduce
rather than a two-way one.  At short sequence lengths the per-socket
GEMM is then too small to cover that fixed cost.  The second reason is
how much there was to recover in the first place.  The Haswell and
Cascade Lake comparison already showed the gain shrinking as cores per
socket rise, because more threads spread the baseline's remote traffic
(Figure~\ref{fig:speedup_bars}b).  With 10 cores per socket against
dual-socket Haswell's 6, the baseline here is less stressed to begin
with.


\begin{figure}[t]
\centering
\includegraphics[width=\columnwidth]{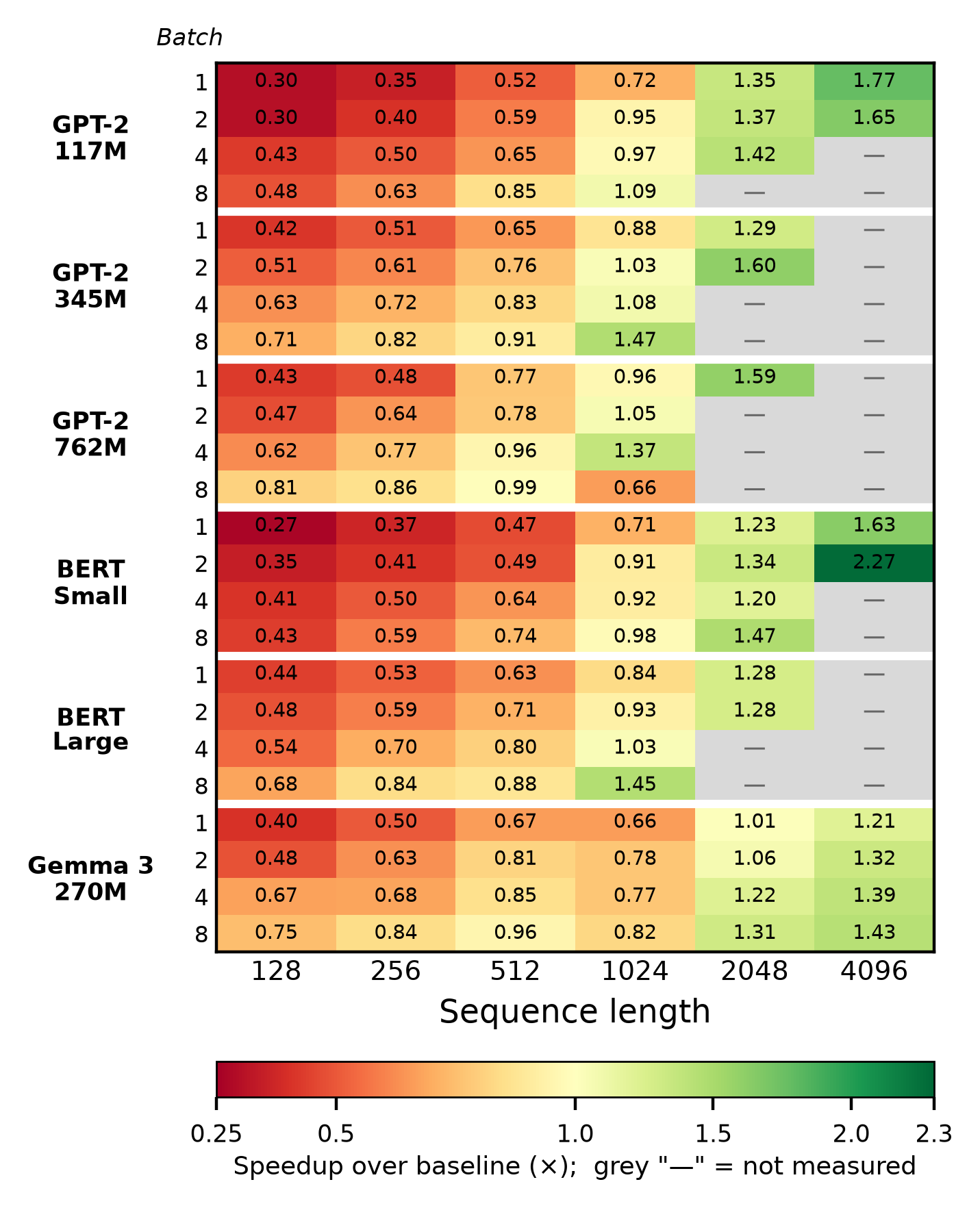}
\caption{End-to-end prefill speedup of \NumaTorch{} over the unmodified
Triton-CPU baseline on the four-socket Haswell-EX platform, for every
evaluated model, batch size and sequence length. Cells are colored about
$1.0\times$, so green cells are faster than the baseline and red cells slower.}
\label{fig:foursocketgrid}
\end{figure}

Figure~\ref{fig:foursocketgrid} reports every individual configuration. Of the
144, the 120 shown are measured; the remaining 24 could not be measured on this
64\,GB host, as the per-socket sharing of activations places their working set
beyond available memory.

\section{Related Work}
\label{sec:related-work}

\subsection{LLM Inference Optimizations on CPUs}
A substantial body of work has sought to alleviate the compute and memory bottlenecks of executing large language models (LLMs) on CPU architectures. Foundational software efforts like \texttt{llama.cpp} deliver high baseline performance for CPU inference through highly optimized, manually engineered C/C++ backends~\cite{gerganov2023llama}. Similarly, specialized frameworks such as ArcLight address the memory bottleneck via cross-NUMA tensor parallelism on many-core CPUs~\cite{xu2026arclight}. However, because these are custom inference engines written entirely in C++, they rely heavily on manual engineering and require meticulous rewriting of model code to support different LLM architectures~\cite{xu2026arclight}. In contrast, our compiler-driven approach integrates seamlessly into the standard TorchInductor pipeline, remaining entirely transparent to the user without requiring modifications to the model code or inference loop.

Algorithmic and hardware-aware optimizations have also been extensively explored to maximize CPU efficiency. Shen et al. proposed an automated INT4 weight-only quantization flow paired with tailored runtime kernels to accelerate CPU deployment~\cite{shen2023efficient}. He et al. introduced KV cache size reduction techniques and distributed inference optimizations relying on the oneAPI Collective Communications Library (CCL)~\cite{he2024inference}. At the architectural level, Dice and Kogan detailed software-level scaling optimizations for Transformer execution on CPUs without degrading model accuracy~\cite{dice2021optimizing}. Furthermore, recent innovations such as NoMAD-Attention replace traditional multiply-add (MAD) computations with ultra-low-latency in-register lookups utilizing CPU Single-Instruction-Multiple-Data (SIMD) capabilities~\cite{nomad2024neurips}.

\subsection{Efficient Attention Mechanisms}
Several of the CPU strategies above target memory directly. Weight-only quantization~\cite{shen2023efficient} and KV-cache compression~\cite{he2024inference} shrink the memory footprint and reduce bandwidth use. These approaches reflect that memory bandwidth, rather than raw compute, is the dominant constraint for LLM inference. IO-aware attention restructures computation around the memory hierarchy. On GPUs, FlashAttention uses aggressive tiling to minimize intermediate transfers between HBM and on-chip SRAM~\cite{dao2022flashattention}. FlashAttention-2 further optimizes thread-block parallelism and work partitioning~\cite{dao2023flashattention2}. Although these algorithms are tailored to the GPU memory hierarchy, they underscore the importance of strict data locality. Our compiler-based methodology applies this principle to multi-socket CPU topologies by minimizing cross-socket interconnect traffic.

\subsection{Distributed Execution and Tensor Parallelism}
To mitigate synchronization overhead and manage massive parameter scales, distributed paradigms like Megatron-LM introduced tensor parallelism to minimize synchronization points within both the Multi-Layer Perceptron (MLP) and attention blocks~\cite{shoeybi2019megatron}. Generalizing distributed tensor computations, Mesh-TensorFlow established a flexible language for compiling Single-Program-Multiple-Data (SPMD) graphs across high-performance supercomputing clusters~\cite{shazeer2018mesh}. Extending these compiler abstractions to modern primitives, Triton-distributed augmented the Triton compiler with OpenSHMEM. This enables native overlapping of computation and communication for distributed AI workloads~\cite{zheng2025triton}. Our methodology adapts large-scale tensor parallelism principles to the PyTorch FX graph. Specifically, it uses column- and row-parallel decompositions within a single multi-socket CPU to localize DRAM access and avoid inter-socket bottlenecks.

Beyond these foundational techniques, production distributed frameworks expose tensor and data parallelism as first-class runtime services. PyTorch's native \texttt{torch.distributed} stack organizes execution as per-device processes that exchange shards through collective primitives. Its APIs include DistributedDataParallel~\cite{li2020pytorch}, Fully Sharded Data Parallel (FSDP)~\cite{zhao2023fsdp}, and a DTensor-based tensor-parallel interface. DeepSpeed combines ZeRO-style state partitioning across ranks~\cite{rasley2020deepspeed,rajbhandari2020zero} with tensor-parallel inference kernels. These systems delegate data movement to collective communication libraries such as NCCL~\cite{nccl}, oneCCL~\cite{oneccl}, Gloo~\cite{gloo}, or MPI. Their \texttt{all-reduce} and \texttt{all-gather} routines are tuned for NVLink, PCIe, or InfiniBand and scheduled to overlap with computation.

Our setting differs along three axes that make these frameworks a poor direct fit for intra-node CPU inference. \emph{Process model:} these frameworks launch one OS process per device. NUMA-TP instead runs in a single process and treats each socket as a rank executed by an affinity-pinned thread group. \emph{Communication:} these frameworks invoke a collective library over a network or device interconnect. By contrast, our cross-socket reduction directly sums socket-local buffers across the QPI/UPI link. \emph{Partitioning interface:} these frameworks expose partitioning through user-facing APIs or model wrapping, often with architecture-specific configuration. NUMA-TP instead uses a compiler FX pass and requires no model-code changes. It recasts distributed tensor parallelism as an intra-node, compiler-level transformation tailored to the low-latency, high-bandwidth cross-socket interconnect.

\subsection{NUMA-Aware Execution Strategies}
Several NUMA-aware deep learning optimizations are close in spirit to our work. We describe their similarities and differences below.

NUMA-Caffe restructures layer-level parallelism into a multi-solver design to improve CNN training throughput on NUMA hardware~\cite{roy2018numacaffe}. Like our work, it treats NUMA locality as important on multi-socket machines. However, its workload and mechanism differ from ours. NUMA-Caffe targets CNN \emph{training} with \emph{data parallelism}, using per-node model replicas over a partitioned batch. It implements this approach as a framework-level Caffe modification. In contrast, we target Transformer \emph{inference} with \emph{tensor parallelism} applied as a compiler transformation. A second line of work characterizes the bottleneck rather than removing it. Studies of cache contention and NUMA remote accesses for Transformer inference on many-core processors~\cite{acm3545022} confirm that remote-DRAM traffic dominates. However, they stop at analysis and do not provide a transparent mechanism to eliminate it.

Closest to our setting is Sandwich~\cite{zhao2025sandwich}, a hardware-centric LLM serving engine. Like NUMA-TP, Sandwich partitions models across NUMA nodes to reduce cross-socket traffic and tailors execution to the prefill and decode phases. The key differences are integration and binding time. Sandwich is a \emph{standalone} engine that partitions models \emph{statically} using ahead-of-time, per-phase execution plans. NUMA-TP instead runs \emph{inside} the standard TorchInductor pipeline as a post-grad FX pass that reuses the Triton-CPU backend. Sandwich, like the C++ engine ArcLight~\cite{xu2026arclight}, requires adopting a separate runtime and porting models to it. NUMA-TP instead dispatches kernels dynamically through a NUMA-aware thread pool and works with unmodified PyTorch model code.


\section{Conclusion}
\label{sec:conclusion}
This paper introduces a compiler-driven, PyTorch-native solution to the
DRAM bandwidth bottleneck of the Large Language Model (LLM) prefill
phase on multi-socket CPUs. Extending the TorchInductor compilation
path, it implements NUMA-aware Tensor Parallelism (NUMA-TP) via
post-grad FX graph transformations, coupled with an extended Triton-CPU
backend for per-socket buffer allocation and NUMA-aware kernel
dispatch. The transformation strictly enforces data locality,
remains entirely transparent to the user, and yields prefill
speedups of up to $2.63\times$ on dual-socket Haswell and $1.87\times$ on Cascade
Lake platforms, exceeding what process-level \texttt{numactl} policies achieve on
the same sweep.

Three directions follow from this work.  First, NUMA-Torch should extend
beyond the transformer sub-layers treated here to other
locality-critical compute blocks, so that any block whose cost is
dominated by remote memory traffic can be partitioned by the same
passes.  Second, the transformation should be chosen rather than applied
unconditionally, with the compiler determining the
communication-computation tradeoff for each block from dataflow analysis
and the characteristics of the target platform, such as its socket count
and interconnect bandwidth.  The four-socket result makes the case for
the second directly, since the same transformation that pays at high
arithmetic intensity does not pay at low.  The third sits on a separate
track within LLM inference optimization.  This work targets prefill, and
carrying NUMA-aware partitioning into decode needs its own treatment of
KV-cache placement and reduction scheduling rather than a direct reuse
of these passes.


The TorchInductor NUMA-TP FX passes and the extended Triton-CPU backend are publicly available at
\url{https://github.com/LLM-Compiler/numa-torch}.

\section*{Acknowledgment}

The authors sincerely acknowledge the financial assistance provided by
the University of Moratuwa to cover the publication fees. Furthermore, the authors extend their gratitude to the anonymous reviewers for their thorough and insightful feedback. 

\bibliographystyle{IEEEtran}
\bibliography{references}

\end{document}